\documentclass[twoside]{article}
\usepackage{fleqn,espcrc2}
\usepackage{graphicx}
\title{Comparison of Hadronic Interaction Models at Auger Energies}

\author{D.~Heck\address{Institut f{\"u}r Kernphysik, Forschungszentrum 
                        Karlsruhe, P.O. Box 3640, D--76021 Karlsruhe, 
        Germany}\thanks{corresponding author, e-mail: dieter.heck@ik.fzk.de},
        M.~Risse$^{\rm a}$, and
        J.~Knapp\address{Dept. of Physics and Astronomy, University of Leeds, 
                         Leeds LS2 9JT, United Kingdom }
       }
\begin{document}

\begin{abstract}
The three hadronic interaction models DPMJET 2.55, QGSJET 01, and SIBYLL 2.1,
implemented in the air shower simulation program CORSIKA, are compared
in the energy range of interest for the Pierre Auger experiment.
The model dependence of relevant quantities in individual hadronic
interactions
and air showers is investigated.
\vspace{1pc}
\vspace{-1.pc}
\end{abstract}

\maketitle

\section{INTRODUCTION}

The engineering array of the southern Pierre Auger Observatory 
(PAO) \cite{auger} in Argentina has recently started to take data.
The interpretation of the measurements requires reliable numerical 
simulations of extensive 
air showers (EAS) in the extremely-high energy (EHE) range $> 10^{19}$~eV.
A large uncertainty in such simulations arises from the models 
which describe the hadronic interactions. 
As one has to extrapolate in energy from accelerators by 
several orders of magnitude and into the forward kinematical range 
which is unobserved by collider experiments, the models have to 
rely on theoretical guidelines to describe the EHE collisions.
In this contribution model predictions for single interactions 
of  p-$\overline{\rm p}$, p-$^{14}$N, and $\pi$-$^{14}$N 
collisions are compared, and we examine how their features 
influence measurable EAS quantities such as the longitudinal 
development with the shower maximum $X_{\rm max}$ 
and the lateral distributions of particle densities at ground.

\section{MODELS}

Presently three hadronic interaction codes coupled with the
EAS simulation program CORSIKA \cite{corsika} are 
able to treat hadronic collisions at energies $>$ 10$^{19}$ eV: 
\begin{figure}[h]
 \begin{center}
  \hspace*{-5pt}
  \includegraphics[width=18.5pc]{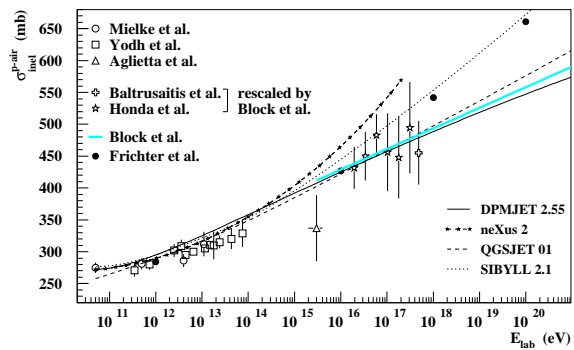}
 \end{center}
 \vspace*{-32pt}
 \caption{Inelastic proton-air cross-sections of the considered models. 
          Included are experimental data of air shower measurements.
          Further details see ref. \cite{knapp-astrop}.}

 \label{fig-sigma_p_air}
 \vspace*{-0.3cm}
\end{figure}
DPMJET 2.55 \cite{ranft}, QGSJET 01 \cite{qgs,hec01}, 
and SIBYLL 2.1 \cite{sibyll,sibyllnew}. 
In this study {\sc neXus 2} \cite{nexus} is partly included despite of its 
upper limit of $< 2 \times 10^{17}~$eV recommended by its authors. 
But already at this energy some interesting trends of this model 
show up.
Fig. \ref{fig-sigma_p_air} shows the proton-air cross-sections 
for production of secondary particles as function of energy.

\section{SINGLE INTERACTIONS}

All models are tuned to reproduce the available collider data. 
They agree fairly well in the pseudo-rapidity, multiplicity, and 
transverse momentum distributions obtained by the UA5 \cite{ua5}, 
CDF \cite{cdf}, and P238 \cite{harr} experiments. 
\begin{figure}[t]
 \begin{center}
  \hspace*{-5pt}
  \includegraphics[width=18.5pc]{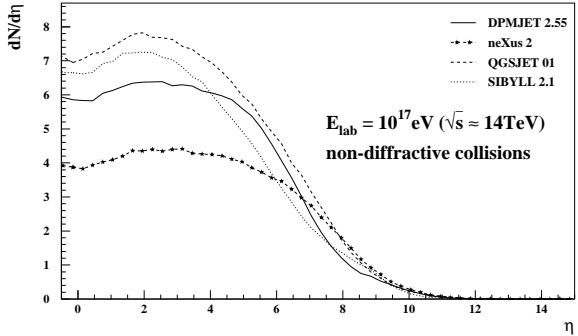}
 \end{center}
 \vspace*{-32pt}
 \caption{Pseudorapidity distribution of charged particles in
          ${\rm p}$-${\overline {\rm p}}$ collisions at 
          $E_{\rm lab}$ = 10$^{17}$ eV.}
 \label{fig-pprap1e17}
 \vspace*{-13pt}
\end{figure}
\begin{figure}[h]
 \vspace{-6pt}
 \begin{center}
  \hspace*{-5pt}
  \includegraphics[width=18.5pc]{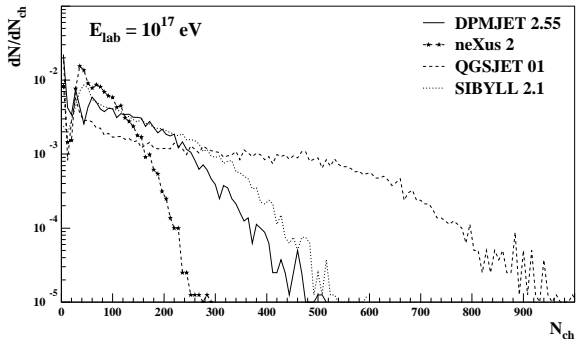}
 \end{center}
 \vspace*{-32pt}
 \caption{Distribution of charged particle multiplicity in
          $\pi$-$^{14}$N collisions at $E_{\rm lab}~$=~10$^{17}$ eV.}
 \label{fig-piNmult1e17}
\end{figure}
But when extrapolating to higher energies,
already at $E_{\rm lab}$ = 10$^{17}~$eV (corresponding to 
$\sqrt{s} \approx 14$ TeV, which will be reached by the future 
LHC-collider), the mid-rapidity density at $|\eta| < 4$  of 
QGSJET exceeds that of {\sc neXus} by up to a factor of 
$\approx$2 as shown in Fig. \ref{fig-pprap1e17}.
In Fig. \ref{fig-piNmult1e17} the charged particle multiplicity 
distribution of  $\pi$-$^{14}$N collisions shows a similar behaviour.
Again QGSJET predictions are higher than those of all other models, 
while the {\sc neXus} 
distribution repeats the low multiplicity already shown in the  
${\rm p}$-${\overline {\rm p}}$ collisions of Fig. \ref{fig-pprap1e17}.

This behaviour reflects the different treatment 
of the Pomeron exchange by the various models. 
In QGSJET a hard Pomeron is always coupled to the partons 
of projectile and target via soft Pomerons, which finally 
produce the large number of secondary particles. 
\begin{figure}[ht]
 \begin{center}
  \hspace*{-7pt}
  \includegraphics[width=18.7pc]{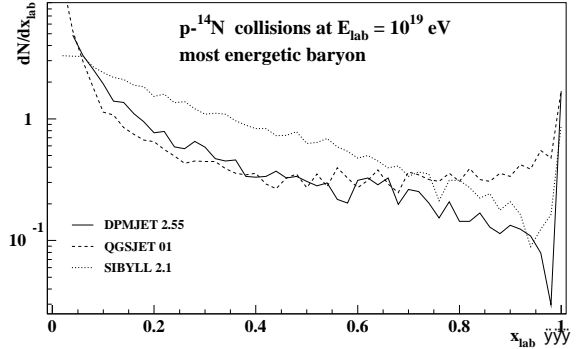}
 \end{center}
 \vspace*{-32pt}
 \caption{Distribution of longitudinal momentum fraction carried away 
          by the most energetic baryon emerging from $\rm p$-$^{14}$N 
          collisions at $E_{\rm lab}~$=~10$^{19}$ eV.}
 \label{fig-pNxf1e19}
 \vspace*{-0.3cm}
\end{figure}
Secondaries produced by cutting soft 
Pomerons appear in the mid-rapidity range 
and take away only a minute energy fraction. 
Therefore they influence the development of charged particle numbers in EAS
rather insignificantly. 

Already in the calculation of the Pomeron exchange probabilities {\sc neXus} 
applies strict energy conservation, which reduces the number of 
cut hard and soft Pomerons, thus reducing the overall multiplicity.
This decreases the number of particles emitted in the 
pseudo-rapidity range below $|\eta| < 6$ while the fraction 
of secondaries produced in the very forward direction 
$|\eta| > 8$ resembles that of the other considered models.

The distribution of the longitudinal momentum fraction taken away by the
leading particle is displayed in Fig. \ref{fig-pNxf1e19}.
Large differences up to a factor of 10 show up in the diffractive
region $x_{\rm lab} > 0.8$, but also around $x_{\rm lab} \approx 0.4$
significant deviations are visible.
The extremely low probability of DPMJET to produce baryons at 
$x_{\rm lab} \approx$ 0.97 in p-$^{14}$N collisions is remarkable. 
It reflects the insufficient knowledge on high- and low-mass 
diffraction: 
Any detailed treatment within the interaction codes is strongly model
dependent.

\section{INFLUENCE ON EAS FEATURES}

In hadronic EAS the $\pi^{\pm}$-mesons are the most frequent secondary 
hadronic 
\begin{figure}[ht]
 \begin{center}
  \includegraphics[width=17.9pc]{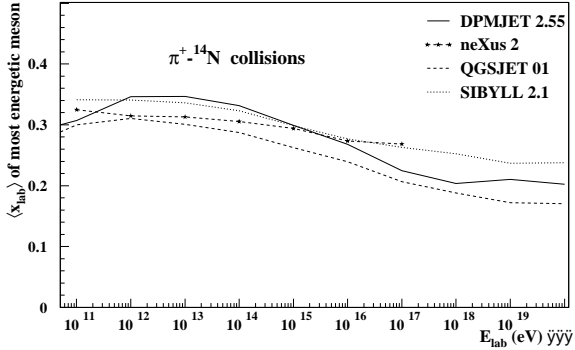}
 \end{center}
 \vspace*{-32pt}
 \caption{Average of longitudinal momentum fraction carried away by 
          the most energetic meson emerging from $\pi$-$^{14}$N 
          collisions as function of energy.}
 \label{fig-piNxaver}
 \vspace*{-13pt}
\end{figure}
particles and nitrogen 
is the most abundant component of air. 
Therefore the character of an EAS is essentially influenced
by the features of $\pi$-$^{14}$N collisions. 
In such interactions QGSJET exhibits a lower average 
elasticity than DPMJET and SIBYLL, see Fig. \ref{fig-piNxaver}. 
Therefore EAS simulated with QGSJET develop faster than those produced with
the other models.

The second important influence on the development process
arises from the cross-sections.
The $\pi$-air production cross-sections behave similar to 
those shown in Fig. \ref{fig-sigma_p_air}, but with 
DPMJET staying well below QGSJET above 10$^{15}$ eV.
Therefore EAS simulated with the DPM\-JET model are expected 
to develop late, while SIBYLL showers will fall somewhere  
\begin{figure}[hb]
 \vspace*{-4pt}
 \begin{center}
  \hspace*{-5pt}
  \includegraphics[width=18.4pc]{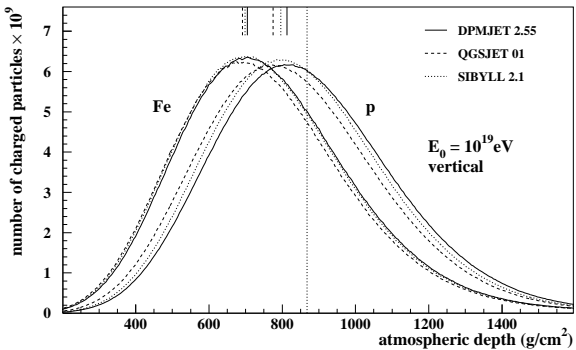}
 \end{center}
 \vspace*{-32pt}
 \caption{Longitudinal development of charged particle number.
          Vertical incidence, $E_0$=10$^{19}~$eV, $E_{\rm e}>0.1~$MeV.
          The vertical dotted line indicates the vertical depth of the PAO
          in Argentina \cite{auger}.}
 \label{fig-longi_charg_1e19}
\end{figure}
\begin{figure}[ht]
 \begin{center}
  \hspace*{-8pt}
  \includegraphics[width=18.5pc]{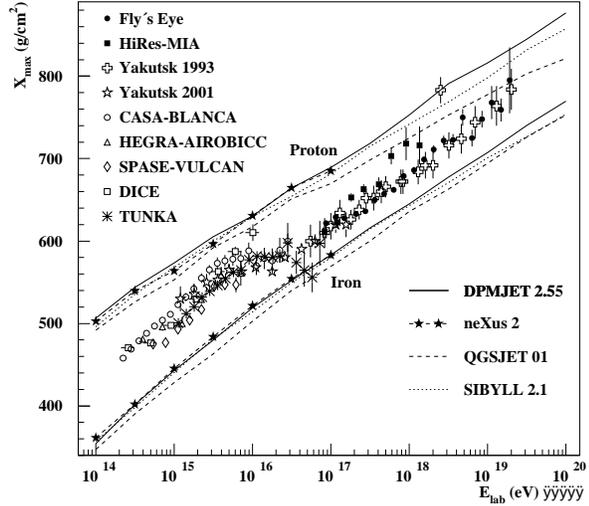}
 \end{center}
 \vspace*{-32pt}
 \caption{Depth X$_{\rm max}$ of shower maximum as function of primary
          energy. Vertical incidence, $E_0$=10$^{19}~$eV,
          $E_{\rm e}>0.1~$MeV.
          References to experiments (fluorescence = full
          symbols, Cherenkov techniques = open symbols) are given in
          \cite{knapp-astrop}.}
 \label{fig-xmax}
 \vspace*{-12pt}
\end{figure}
in between QGSJET and DPMJET. 
This is demonstrated in Fig. \ref{fig-longi_charg_1e19} 
for proton and Iron induced showers of 10$^{19}$ eV.
At the upper margin of Fig. \ref{fig-longi_charg_1e19}
the positions of the maxima are indicated. They nearly coincide for 
Iron induced showers, but for proton induced EAS the difference
between QGSJET and DPMJET amounts to $\approx$ 40 g/cm$^2$.

The depth of shower maximum $X_{\rm max}$ as function of 
primary energy is plotted in Fig.~\ref{fig-xmax}. 
While for Iron induced showers even at highest energies 
the $X_{\rm max}$ values coincide within $<~20~$g/cm$^2$,
with a tendency of DPMJET to predict deeper pen\-e\-tration, 
the lines of proton induced showers show a clear divergence
with increasing energy reaching 55~g/cm$^2$ at 10$^{20}~$eV. 
Especially the QGSJET distribution flattens with increasing energy.
Starting at 10$^{15}$ eV with 65~g/cm$^2$ 
per energy decade the slope reaches 45~g/cm$^2$ per decade 
at 10$^{20}$~eV. 
The flattening has to be attributed to the lowering of the
$\pi$-$^{14}$N 
\begin{figure}[ht]
 \begin{center}
  \hspace*{-7pt}
  \includegraphics[width=18.6pc]{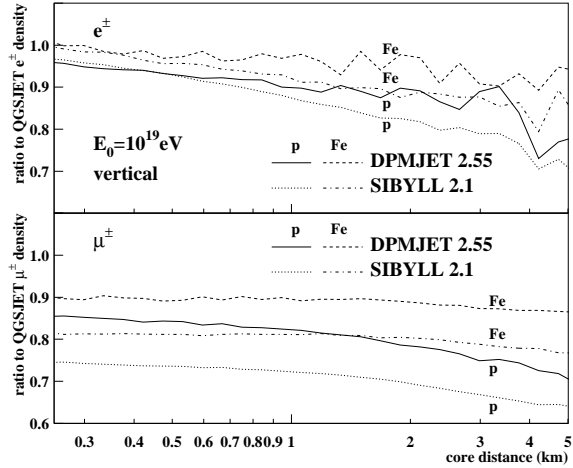}
 \end{center}
 \vspace*{-32pt}
 \caption{Ratios of lateral particle densities relative to QGSJET.
          Vertical incidence, $E_0$=10$^{19}$~eV, 
          $\varepsilon_{\rm thin}$ = 10$^{-6}$ (optimized \cite{ris01}), 
          $E_{\rm e} > 0.25$ MeV, $E_{\mu}>0.1$~GeV.}
 \label{fig-lat_1e19_e_mu_ratio}
 \vspace*{-10pt}
\end{figure}
elasticity (Fig. \ref{fig-piNxaver}) 
together with an increasing $\pi$-$^{14}$N cross-section  
of QGSJET which exceeds that of DPMJET by 10\% at 
$E_{\rm lab} = 10^{17}~$eV.
Above 10$^{15}$ eV the $X_{\rm max}$ slopes of the other 
models show no significant change for proton induced EAS.

Because of its early development QGSJET exhibits 
the flattest lateral distribution for electrons and for muons.
This results in the highest particle densities
at distances $>300$ m
not only for proton but also for Iron induced EAS.
Therefore we compare in Fig. \ref{fig-lat_1e19_e_mu_ratio} 
the lateral distributions of the other models relative 
to those of QGSJET. 
With increasing distance from the shower core the 
densities predicted by the other models are always smaller. 
The effect is more pronounced for proton induced showers.
In general the SIBYLL densities are lowest and the
differences are larger for the muon distributions.

\section{CONCLUSIONS}

We emphasize the importance of future collider experiments 
to measure the very forward range of particle production thus 
reducing the present uncertainties in EAS simulations 
caused by the hadronic interaction models. 
Despite the unknown elemental composition of the primary cosmic 
radiation at EHE, the PAO should be able to 
constrain the hadronic interaction models by evaluating hybrid EAS 
events which combine the longitudinal information measured by  
fluorescence telescopes and the lateral particle densities 
registered by water Cherenkov surface detectors. 

\section*{Acknowledgments}
Many thanks go to the authors of the hadronic interaction models for 
their advice in linking their codes with CORSIKA. 
We thank S. Ostapchenko for enlightening discussions and 
R. Engel for carefully reading the manuscript.
The authors acknowledge support for a British-German Academic Research
Collaboration from The British Council and the DAAD.

\end{document}